# Multivariable functions approximation using a single quantum neuron

A.A. Ezhov[1,2], A.G. Khromov[2], and G.P. Berman[1]

[1] Theoretical Division and CNLS, Los Alamos National Laboratory,
Los Alamos, New Mexico 87545

[2] Troitsk Institute of Innovation and Fusion Research
142092, Troitsk, Moscow region, Russia

**ABSTRACT**

We describe a model element able to perform universal stochastic approximations of continuous multivariable functions in both neuron-like and quantum form. The implementation of this model in the form of a multi-barrier, multiple-slit system is proposed and it is demonstrated that this single neuron-like model is able to perform the XOR function unrealizable with single classical neuron. For the simplified waveguide variant of this model it is proved for different interfering quantum alternatives with no correlated adjustable parameters, that the system can approximate any continuous function of many variables. This theorem is applied to the 2-input quantum neural model based on the use of the schemes developed for controlled nonlinear multiphoton absorption of light by quantum systems. The relation between the field of quantum neural computing and quantum control is discussed.

## I. Introduction

Current attempts to combine quantum and neural information processing can be justified both by the potential of combining their benefits and also to eliminate some their inherent shortcomings [1].

It is well known, that the strong demand in neural and quantum computations is driven by the limitations in the hardware implementation of classical computations. Classical computers efficiently operate with *numbers* (integer and real) and *symbols*, processing relatively short bit registers $d < 128$. But the processing of *patterns* (wide-band signals having $d >> 100$ bits) is limited by the empirical Rent law which demands in this case the use of enormous number of gates $\propto d^{4.8}$ to process $d$ bit registers. This motivates the search for new architectures able to process wide bandwidth. Moreover, the typical computer program able to perform universal calculations on patterns requires $\propto 2^d$ operators [2]. This fact completely excludes the possibility of using an *algorithmic approach*.

Artificial neural networks (ANN) give answers to both of these challenges, suggesting the use of novel architectures able to process long bit strings, not using few-bit gates, and learning by example, not programming. ANN can solve complex problems typical for poorly formalized knowledge domains (*theory-poor* but *data-rich* applications). ANN also have other attractive features including parallel distributed processing and robustness.

Quantum computations also have their historical principal roots in hardware limitations, associated with the miniaturization of computer elements, which will be governed by quantum laws. The mainstream of research in quantum computing deals with the development of the quantum analog of *classical* computational architectures which operate with qubits strings using few-qubit gates in sequential operations for which precisely formulated *algorithms* must be used. Algorithms of P. Shor [3] and L. Grover [4] are examples.

Quantum computers retain many features inherent in classical computers. They cannot operate with wide-band signals and cannot be simply trained by examples. Their efficiency will depend on the discovery of sophisticated and powerful quantum algorithms.

Classical neural networks also face many difficulties, including the absence of rules for determining optimal architectures, their time-consuming training, and their restricted memory capacity (in content-addressable memory models).

The quantum approach seems to be useful in overcoming at least some of these difficulties. For example, as shown by D. Ventura [5], *quantum associative memory* can has exponential capacity, while T. Menneer [6] has argued that quantum superposition of the outputs of many networks permits the use of simpler and faster trainable architectures operated in parallel universes [7].

It is important to take into account one principal difference between classical and neural computations (and also between computers and neurocomputers). While classical computers process digital information, the neurocomputers are inherently analog. Despite the historical use of neural networks for logical functions, their modern applications mainly deal with analog input and output. Consequently, the concept of a qubit can be as irrelevant to future quantum neural technology as the concept of bit is to modern neural technology. There is also a quantum computing approach to the information processing of continuous variables [8,9]. It uses the fact that some quantum variables (position, momentum etc.) have a continuous spectrum. Here we shall use a different approach to quantum neural models. The output of most the widely applicable neural systems – multilayer perceptrons – can be interpreted as the *a posteriori* probability for the input to belong to a given class. Therefore, it is reasonable to propose that quantum mechanical probabilities can be used to generate the output of quantum neural systems, enabling stochastic calculations with many particles.

Three main theoretical results form the basis of neural technology processing analog information.

1. The proof that the 2-layer perceptron is the universal approximator of continuous multivariable functions [10,11]. These systems can be used to solve an enormous number of problems in pattern classification, categorization, regression, compression, etc.
2. The discovery of the surprisingly efficient method for training of multilayer perceptrons (*back propagation error* method [12]). These systems *can be trained* even using personal computers.
3. The proof that multilayer perceptrons have finite Vapnik-Chervonenkis dimensionality [13]. Consequently, these neural architectures can *generalize* from data.

Despite the long history of neural networks research, the first result was obtained only in 1989 [10,11] and the second one became widely known in 1986 [12]. Surely, analogous results will be extremely important for the quantum analog of classical neural systems.

Here we show, that for simplified quantum neural systems, the analog of point 1 can be proved. This result can be extended to other variants of quantum neural processing. Surprisingly the use of the quantum approach can eliminate the necessity of building a *networks* of neurons to obtain approximate universality. *Only* quantum neuron seems to be able to perform universal approximations, the way a single photon can "investigate" all possible paths connecting initial and final positions.

## II. Multi-barrier multiple-slit model

First we define a system to be "neural" if there is at least one "neuron" in it. It is possible to call a unit neuronal if:

- it has many (*d*) inputs and single output;
- the external stimulus $\mathbf{x} = (x_1, \ldots, x_d)$ is weighted by a synaptic vector $\mathbf{w}$ and
- the resulting neural activity, $a = \mathbf{w}^T \mathbf{x}$, is transformed *nonlinearly* into the unit's output $y = f(a)$.

These properties are sufficient for the unit to be a "neuron".

We can claim that a *neural* system is a *quantum neural* system if it can perform a *quantum computation*. Here we use the quantum computation defined by A. Narayanan [14]. The main properties of this algorithm are:

- the problem to be decided can be split in subproblems;
- each subproblem is decided in separate "universe";
- the output involves the interference of different "universes".

Note, that this definition of quantum algorithm is based on the use of Everett's many-universes interpretation of quantum mechanics[1] [15].

A quantum neural system is physically *implementable* if one can demonstrate how it can be realized as a physical quantum device.

Consider typical problem in classical neurocomputing. It is necessary to construct a neural system able to perform the mapping from the set of input patterns to the set of prescribed real values (regression) $\mathbf{x}^\alpha \to y^\alpha$, $\mathbf{x}^\alpha = (x_1^\alpha, \ldots, x_d^\alpha)$ $\alpha = 1, \ldots, P$.

Consider a system of $d-1$ barriers with multiple slits analogous to the one introduced by R. Chrisley [16] (Fig. 1). In contrast to his model suppose, that the space between barriers can be filled with substances having different refractory indexes $n_j, j = 1, \ldots, d$. The vector $\mathbf{n} = (n_1, \ldots, n_d)$ will serve as the *input* to our system, analogous to the vector $\mathbf{x}$.

Let $l(s_{k_j}^j, s_{k_{j+1}}^{j+1})$ denote the path length between two slits $s_{k_j}^j, s_{k_{j+1}}^{j+1}$ in the *j*-th and (*j+1*)-th barriers $k_j = 1, \ldots, N_j$. $N_j$ = the number of slits for *j*-th barrier. The values of *l* will serve as adjustable parameters proportional to the weight coefficients, $w_i$.

Suppose detector (D) has a definite location on the screen. Let the complex amplitude for the particle's source-detector transition along a *given trajectory* $S \to s_{k_1}^1 \to \ldots s_{k_{d-1}}^{d-1} \to D$ serve as the output of the system for a given IA: $y = \langle D | s_{k_{d-1}}^{d-1} | \ldots s_{k_1}^1 | S \rangle$. Note, that it is convenient to consider not *l*, but the positions of the slits $r_{k_j}^j$ as trainable parameters. Then

$$l(s_{k_j}^j, s_{k_{j+1}}^{j+1}) = \sqrt{h_j^2 + (r_{k_j}^j - r_{k_{j+1}}^{j+1})^2},$$

where $h_j$ is the distance between *j*-th and (*j+1*)-th barriers.

We now calculate the probability amplitude for the particle to reach the detector by the trajectory

$$S \to s_{k_1}^1 \to \ldots \to s_{k_{d-1}}^{d-1} \to D$$

$$\langle D | s_{k_{d-1}}^{d-1} \rangle \ldots \langle s_{k_1}^1 | S \rangle = \frac{1}{\prod_{j=1}^d l(s_{k_{j-1}}^{j-1}, s_{k_j}^j)} \times \exp\left(\frac{2\pi i}{\lambda} \sum_{j=1}^d n_j l(s_{k_{j-1}}^{j-1}, s_{k_j}^j)\right), \quad (1)$$

where $s_{k_0}^0 \equiv S$, $s_{k_d}^d \equiv D$.

$$\text{Define} \quad \theta = \ln\left(\prod_{j=1}^d l(s_{k_{j-1}}^{j-1}, s_{k_j}^j)\right), \quad (2)$$

$$w_j = \frac{2\pi}{\lambda} l(s_{k_{j-1}}^{j-1}, s_{k_j}^j), \quad j = 1, \ldots, d. \quad (3)$$

Then amplitude (1) can be rewritten as

$$\langle D | s_{k_{d-1}}^{d-1} \rangle \ldots \langle s_{k_1}^1 | S \rangle = \exp(\mathbf{wn} - \theta). \quad (4)$$

We conclude that the calculation of the amplitude corresponding to given trajectory joining the source of the particle and the detector satisfies to our earlier definition of computation performed by *single neuron*. Therefore, our experimental system has *single neuron in each universe* (IA). The full amplitude can be obtained by summation over all these trajectories

$$\langle D|S\rangle = \sum_{\{\mathbf{k}\}} \langle D|s^{d-1}_{k_{d-1}}\rangle...\langle s^1_{k_1}|S\rangle. \tag{5}$$

---

[1] We shall also use the abbreviation IA (Interfering Alternatives) for the interfering universes.

If we define the output of our problem to be the probability of the particle's detection $|\langle D|S\rangle|^2$, then this value can be considered as the result of interference of the outputs of the quantum neurons. Our simple system really has the main properties of both neural and quantum computations. Let us consider an example in which this system is used for solving a famous neurocomputing problem.

Consider the classical Young single-barrier double slit We can introduce in both parts of the system (before and after the barrier) glasses with refractivity index 1 or *n*. Hence, we can present 4 variants of the input to the system: (1,1), (1, *n*), (*n*, 1), and (*n*, *n*). The values 1 and n can be considered as analogs of binary input variables.

Let us adjust the parameters of this system so that output of the system will produce the value of the XOR function:

| $x_1$ | $x_2$ | Y |
|---|---|---|
| 1 | 1 | 0 |
| 1 | n | 1 |
| n | 1 | 1 |
| n | n | 0 |

Choose $n=5/3$, $h_0 = h_1 = h$. $r^{(2)}(D) = 0$, and suppose that $\lambda \ll h$.

Let us also choose locations of splits in such a way that path lengths of particle in both parts of trajectory (before and after the barrier) in path 1 ($l_1^1, l_1^2$) exceeds those in path 2 ($l_2^1, l_2^2$) by $3/4\lambda$ (Fig. 2). This can be done choosing

$$r_1 = r_2 + \sqrt{1+h^2/r^2} \cdot 3\lambda/4.$$

For these parameters we obtain:

$$\langle D|s_1^1\rangle\langle s_1^1|S\rangle = \frac{1}{l_1^1 l_1^2}\exp\left(\frac{2\pi i}{\lambda}(n_1 l_1^1 + n_2 l_1^2)\right), \langle D|s_2^1\rangle\langle s_2^1|S\rangle = \frac{1}{l_2^1 l_2^2}\exp\left(\frac{2\pi i}{\lambda}(n_1 l_2^1 + n_2 l_2^2)\right)$$

Remembering that $\lambda \ll h$ we can approximately set that the probability for the particle to be detected (which is the result of interference of the outputs of two neurons) to be:

$$P(n_1, n_2) = |\langle D|s_1^1\rangle\langle s_1^1|S\rangle + \langle D|s_2^1\rangle\langle s_2^1|S\rangle|^2 \cong \frac{1}{(r^2+h^2)^2}|1+\cos\frac{2\pi}{\lambda}(n_1+n_2)\Delta l|^2 \tag{6}$$

where $\Delta l = l_1^1 + l_1^2 - l_2^1 - l_2^2 = 3\lambda/2$.

Then, from (6) it immediately follows that

$$P(1,1) = A|1+\cos 3\pi|^2 = 0$$
$$P(1,n) = A|1+\cos 4\pi|^2 = 4A$$
$$P(n,1) = A|1+\cos 4\pi|^2 = 4A$$
$$P(n,n) = A|1+\cos 5\pi|^2 = 0$$

Hence, the output of the system coincides with the value of XOR with the accuracy of the normalizing factor $4A = 4(r^2 + h^2)^{-2}$. Note, that a single quantum neuron is able to realize XOR function. Classical approach requires 2-layers network of analogous simple neurons.

In general, training of this system can be performed using different optimization procedures. We used a variant of simulated annealing to realize all Boolean functions of two variables. Briefly, the following operators describe the corresponding scheme:

*Simulated Annealing*:
Define the vector of adjustable variables
$$\mathbf{w} = (\{s_{k_j}^j\}, j = 0,...,d; k_j = 1,...,N_j; \{h_j\}, j = 1,...,d-1)$$

To minimize the error functional $E(\mathbf{w}) = \sum_{\alpha=1}^{\aleph}(P(\mathbf{n}^{(\alpha)}, \mathbf{w}) - t^{(\alpha)})^2$, where $\{\mathbf{n}^{(\alpha)} \mapsto t^{(a)}\}, \alpha = 1,...,\aleph$ given set of teaching examples (associations between vectors of refractory indexes and probability for the photon to strike the detector) do the following

    initialize $\mathbf{w} = \mathbf{w}^{(0)}$ as a random vector;
    set the initial temperature $T(0) = T^0$, and the number of steps for
    temperature updating $\tau_{max}$;
    set the step counter $\tau = 0$;
    **until** vector of adjustable parameters become stable **do**:
        $\tau = \tau + 1$;
        randomly increment adjustable parameters $w_i = w_i + \Delta w_i$
        choose sign of $\Delta w_i$ at random;
        Calculate the error change $\Delta E = E(\mathbf{w} + \Delta \mathbf{w}) - E(\mathbf{w})$;
        Update the vector $\mathbf{w}$ according to the probabilistic rule:
$$\Pr(w \to w + \Delta w) = \begin{cases} 1, & \Delta E < 0 \\ \exp(-\Delta E / T(t)), & \Delta E > 0 \end{cases}$$

    **if** $\tau = \tau_{max}$
      update temperature $T(t) = \beta E(t) \exp(-\gamma t)$ (where $\beta$ and $\gamma$ are real positive
      parameters of adaptation);
      set $\tau = 0$.

Note, that the realization of Boolean functions is far from the central problem of really applicable quantum neural technology. But it permits us, at least qualitatively, to study the generalization abilities of these a systems. It can be seen from Fig. 3, that a good generalization (expressed in a smooth mapping performed by the neuron at intermediate values of refractory indexes) is achieved for small difference in optical lengths of photons with different interfered paths ($\Delta l \approx \lambda / 2\Delta n$). Otherwise, corresponding mapping has a non-regular oscillating form.

## III. Waveguide model

It is not easy to prove the universality of the scheme described in previous section. The reason is that photon paths with different trajectories are geometrically correlated. But if we shall neglect these correlations then the required proof can be done. Actually, this simplification means that we switch to a model in which the slits are connected by optical waveguides having prescribed refractory indexes and *arbitrary lengths*. What is more, we shall suppose, that two any slits can be connected by *arbitrary number* of identical waveguides. We shall refer to this model as the *waveguide* –WGM (Fig. 4).

Let us prove that WGM is the universal stochastic approximator of any continuous function. Below we shall use the following notations.

We denote by **R** the field of real numbers and by **C** the field of complex numbers. The sign ■ will denote the end of the proof.

***Lemma***: For each $L \in \Im$, where $\Im$ denotes some set of indexes, let $f_L$ and $g_L$ be complex-valued continuous functions defined on a compact (closed and bounded) subset $E$ of Euclidian space $\mathbf{R}^n$. For any pair of indexes $L, L_1 \in \Im$, let
$$|f_L(\mathbf{x})| = |f_{L_1}(\mathbf{x})| \quad \text{for all } \mathbf{x} \in E. \tag{7}$$
Suppose, that for any real $\varepsilon > 0$, there exists an $L \in \Im$, such that
$$|f_L(\mathbf{x}) - g_L(\mathbf{x})| < \varepsilon \quad \text{for all } \mathbf{x} \in E.$$
Then, for any real $\varepsilon > 0$ there exists an $L \in \Im$, such that
$$||f_L(\mathbf{x})|^2 - |g_L(\mathbf{x})|^2| < \varepsilon \quad \text{for all } \mathbf{x} \in E.$$

***Proof***: Let $\overline{C}(0, R)$ be a closed ring with radius R and with center at the origin of the complex plane **C** (e.g. the set of all complex numbers z, for which $|z| \leq R$). Because E is compact and $f_L$ is continuous, then the set $f_L(E)$ is compact, and, bounded and belongs to some closed ring $\overline{C}(0, R)$. According to (7) this ring is common for all $L \in \Im$, i.e. $f_L(\mathbf{x}) \in \overline{C}(0, R)$ for all $L \in \Im$ and for all $\mathbf{x} \in E$.

Given $\delta > 0$ the complex function $z \mapsto |z|^2$ is uniformly continuous inside the closed ring $\overline{C}(0, R+\delta)$, e.g. for any $\varepsilon > 0$ there exists a $\delta_1 > 0$, such that for all $z, z_1 \in \overline{C}(0, R+\delta)$ from $|z - z_1| < \delta_1$ it follows that $||z|^2 - |z_1|^2| < \varepsilon$.

Define $\delta_2 = \min(\delta, \delta_1)$ and choose $L \in \Im$ such that $|f_L(\mathbf{x}) - g_L(\mathbf{x})| < \delta_2$ for all $\mathbf{x} \in E$. Then, for any $\mathbf{x} \in E$ from $f_L(\mathbf{x}) \in \overline{C}(0, R)$ it follows that $g_L(\mathbf{x}) \in \overline{C}(0, R+\delta)$. It means that the complex numbers $f_L(\mathbf{x})$ and $g_L(\mathbf{x})$ are both inside the ring $\overline{C}(0, R+\delta)$. Apart from this, they satisfy $|f_L(\mathbf{x}) - g_L(\mathbf{x})| < \delta_2 \leq \delta_1$. Therefore, $||f_L(\mathbf{x})|^2 - |g_L(\mathbf{x})|^2| < \varepsilon$ ■

***Theorem***: Let $f$ be a real non-negative function, defined on a compact subset $E$ of Euclidian space $\mathbf{R}^n$ and let $\omega$ be positive real number. Then, for any $\varepsilon > 0$ there exists an integer $U$ and

(i) *integer* positive digits $k_u$, $u = 1,...,U$;

(ii) real positive digits $l_u^1,...,l_u^d$, $u = 1,...,U$

such that the complex-valued function

$$g(x_1,...,x_d) = \sum_{u=1}^{U} \frac{k_u}{l_u^1 \cdot ... \cdot l_u^d \theta_u} \exp\{i\omega(l_u^1 x_1 + ... + l_u^d x_d + \theta_u)\} \qquad (8)$$

satisfies

$$|f(\mathbf{x}) - |g(\mathbf{x})|^2| < \varepsilon \quad \text{for all } x = (x_1,...,x_d) \in E \qquad (9)$$

***Proof:*** Because the function $\sqrt{f}$ is continuous on E, then for every $\varepsilon > 0$ there exists an integer U and

(i') real positive $A_u$, $u = 1,...,U$,

(i'') real $l_u^1,...,l_u^d, \theta_u$, $u = 1,...,U$ ($\theta_u$ is the analog of bias value),

such that the function

$$g_0(\mathbf{x}) = \sum_{u=1}^{U} A_u \exp\{i\omega(l_u^1 x_1 + ... + l_u^d x_d + \theta_u)\} \qquad (10)$$

satisfies $|\sqrt{f(\mathbf{x})} - g_0(\mathbf{x})| < \varepsilon$ for all $x = (x_1,...,x_d) \in E$. This is an obvious consequence of the Stone-Weierstrass theorem [17]. If E is rectangular it follows from the elementary theory of Fourier series.

Let us choose $\varepsilon > 0$ and apply the lemma. As functions $g_L$ we take the function $g_0$ (10) satisfying the conditions (i') and (i''), by labeling them with some set of indexes $\mathfrak{J}$, and by taking $f_L = \sqrt{f}$ for all $L \in \mathfrak{J}$. We conclude, that there exists a function $g_0$ for which

$$|f(\mathbf{x}) - |g_0(\mathbf{x})|^2| < \varepsilon/2 \quad \text{for all } x = (x_1,...,x_d) \in E. \qquad (11)$$

Now we replace the function $g_0$ by the function having the same form (10) which satisfies the conditions (i') and (11), but for instead to (i'') the more strong condition (ii) will be satisfied.

For this, choose a real positive number $l$ such as

$$-l < l_u^i \quad \text{for all } u = 1,...,U; \quad i = 1,...,d,$$

and real positive number $\theta$ such as

$$-\theta < \theta_u \quad \text{for all } u = 1,...,U.$$

Then the function $g_1(\mathbf{x}) = \exp\{i\omega(lx_1 + ... + lx_d + \theta)\}g_0(\mathbf{x})$ will have the same modulus as $g_0(\mathbf{x})$: $|g_1(\mathbf{x})| = |g_0(\mathbf{x})|$. Therefore, from (11),

$$|f(\mathbf{x}) - |g_1(\mathbf{x})|^2| < \varepsilon/2 \text{ for all } \mathbf{x} \in E.$$

The function $g_1(\mathbf{x})$ can be written in the same form as the function $g_0(\mathbf{x})$ in (10)

$$g_1(\mathbf{x}) = \sum_{u=1}^{U} A_u \exp\{i\omega[(l + l_u^1)x_1 + ... + (l + l_u^d)x_d + (\theta + \theta_u)]\}.$$

Here we have $l + l_u^i > 0$ for all $u = 1,...,U; i = 1,...,d$ and $\theta + \theta_u > 0$ for all $u = 1,...,U$. Hence, we can suggest, that function $g_0(\mathbf{x})$ in (10) satisfies conditions (i'), (ii) and (11).

For arbitrary real $L > 0$ consider the function $g_{0L}(\mathbf{x}) = \exp(i\omega L)g_0(\mathbf{x})$. For any $L > 0$ we have $|g_{0L}(\mathbf{x})| = |g_0(\mathbf{x})|$. Hence, according to (11),

$$|f(\mathbf{x})-|g_{0L}(\mathbf{x})|^2|<\varepsilon/2 \text{ for all } \mathbf{x}\in E. \tag{12}$$

The function $g_{0L}(\mathbf{x})$ can be also written in the form

$$g_{0L}(\mathbf{x})=\sum_{u=1}^{U}\frac{a_u(L)}{l_u^1\cdot...\cdot l_u^d(L+\theta_u)}\exp\{i\omega[l_u^1 x_1+...+l_u^d x_d+(L+\theta_u)]\},$$

where coefficients $a_u(L)>0$ for $u=1,...,U$ depend on $L$.

For every real $L>0$ consider also the function

$$g_L(\mathbf{x})=\sum_{u=1}^{U}\frac{k_u(L)}{l_u^1\cdot...\cdot l_u^d(L+\theta_u)}\exp\{i\omega[l_u^1 x_1+...+l_u^d x_d+(L+\theta_u)]\}$$

where $k_u(L)=[a_u(L)]+1$, $u=1,...,U$. Here $[a_u(L)]$ denotes integer part of $a_u(L)$, i.e. maximal integer lower than $a_u(L)$. Note, that $k_u(L)$ $u=1,...,U$ are positive integer numbers for all $L>0$.

For every $L>0$, we have

$$|g_L(\mathbf{x})-g_{0L}(\mathbf{x})|\le\sum_{u=1}^{U}\frac{1}{l_u^1\cdot...\cdot l_u^d(L+\theta_u)}\le\frac{1}{L}\sum_{u=1}^{U}\frac{1}{l_u^1\cdot...\cdot l_u^d}$$

for all $\mathbf{x}\in E$. Hence, for any $\varepsilon>0$ and sufficiently large $L$ we obtain

$$|g_L(\mathbf{x})-g_{0L}(\mathbf{x})|<\varepsilon$$

for all $\mathbf{x}\in E$. Let us apply the Lemma, by taking $g_{0L}$ as $f_L$. We obtain, that for $\varepsilon$ chosen above (in (11) and (12)) there exists such $L=L^*$, for that

$$||g_{L^*}(\mathbf{x})|^2-|g_{0L^*}(\mathbf{x})|^2|<\varepsilon/2$$

for all $\mathbf{x}\in E$. Together with the inequality (12) which holds, in particular, for $L=L^*$ this gives

$$|f(\mathbf{x})-|g_{L^*}(\mathbf{x})|^2|<\varepsilon$$

for all $\mathbf{x}\in E$. Hence, function $g=g_{L^*}$ satisfies (9) and has the form (8) satisfying also (i) and (ii). ∎

*Corollary*: As it can be seen from the proof, for given $\varepsilon$ the values of $\theta_u$ can be chosen arbitrary large by enlarging only $k_u$ and by holding $l_u^1,...,l_u^d$, $u=1,...,U$ fixed.

One additional remark is needed. As it is seen from this proof, we assume that one additional barrier (waveguide junction node plane) is used in order to have threshold-like adjustable parameters. In the context of previous multi-barrier, multiple-slit model, this requires the presence of some additional inter barrier region filled by the media with fixed refractory index, $n_0\equiv 1$. This setting is also common for classical neural models.

**IV. Multiphoton absorption model**

Note, that we proved the theorem for the waveguide model, leaving intact the original multi-barrier, multiple-slit scheme. Remember, that the complexity of the last one is due to the correlations between amplitudes and phases of photons with different paths. However, if it is possible to perform independent control of both these parameters then we shall have a system for which universality of approximation can be easily derived from the theorem proved above. As one such possibility, let us consider the realization of a quantum neuron having 2 inputs, which is free from correlations of the adjustable parameters describing interference.

The corresponding scheme is in some sense analogous to the coherent quantum control of multiphoton transitions using shaped ultrashort optical pulses, which can be obtained, for example, with proper phase tuning. This approach has been recently proposed by Meshulach and Silberberg [18]. Our approach differs from their scheme in that the amplitude tuning is also used.

In coherent quantum control a quantum system (atom, molecule etc.) should occupy the desired excited state when appropriately stimulated by light. If, for example, a laser pulse has a finite spectral width and the central frequency is only half of the frequency of transition of the chosen quantum system from its ground ($g$) to the excited ($e$) states, then this transition can be only due to two-photon absorption which can be realized in many interfering ways [18]. Using second-order perturbation theory it can be shown that in the case of non-resonant interaction this transition probability can be expressed as [18]

$$P_{g \to e}^{2-PH} \propto \left| \int_{-\infty}^{\infty} d\Omega A(\omega_0/2+\Omega) A(\omega_0/2-\Omega) \exp[i(\Phi(\omega_0/2+\Omega)+\Phi(\omega_0/2+\Omega))] \right|^2. \quad (13)$$

Here $A(\omega)$ and $\Phi(\omega)$ are the spectral amplitude and the spectral phase distributions respectively, and $\omega_0 = (E_f - E_g)/\hbar$ is the transition frequency. The integral on right side of (13) describes the interference of all alternatives corresponding to the absorption of different pairs of photons, whose frequencies $\omega_1$ and $\omega_2$ sum to $\omega_0$. Because one photon has a frequency *lower* than $\omega_0/2$ and the other has *greater* than $\omega_0/2$ and these two frequencies are spatially separated by a diffraction grating in the Meshulach and Silberberg setup (see Fig. 5) such that both photons take paths lower and upper than the center of the programmable one-dimensional liquid-crystal spatial light modulator (SLM) array, then it is possible to use factorized phase tuning so that all phase increments of the SLM cells over the center will be multiplied by an *input* component $x_1$, while the other one by other input component $x_2$. There can exist different ways to achieve this factoring. For example, we can consider $x_1$ and $x_2$ as factors of the electrical fields, which govern the refractory indexes of the SLM cells. In this case we can express the phase increments of both photons as $\Delta\Phi(\omega_1)x_1$ and $\Delta\Phi(\omega_2)x_2$. Here, the multiplication factors $\Delta\Phi(\omega_1)$ and $\Delta\Phi(\omega_2)$ can be set by programmable liquid-crystal spatial light modulators. They should be considered as adaptive parameters of the scheme. Moreover, SLM is also able to independently adjust the spectral amplitudes, $A(\omega)$. Therefore, the probability of the target quantum system excitation due to the two-photon absorption will be of the form

$$P_{g \to e}^{2-PH} \propto \left| \int_{-\infty}^{\infty} d\Omega a(\Omega) \exp[i(\theta(\Omega)+w_1(\Omega)x_1+w_2(\Omega)x_2)] \right|^2, \quad (14)$$

here $a(\Omega) = A(\omega_0/2+\Omega)A(\omega_0/2-\Omega)$; $\theta(\Omega) = \Phi_0(\omega_0/2-\Omega)+\Phi_0(\omega_0/2+\Omega)$ is a sum of the spectral phases of the pulse before SLM, $w_k(\Omega) = \Delta\Phi(\omega_0/2+\Omega \cdot sign(k-3/2)), k=1,2$.

It is easy to see that the universality of this 2-input neuron follows directly from the previously proved theorem if we shall take into account that:

1) the integral on the right side of (14) should be calculated over a narrow phase interval $\Omega \in [-\Omega_0, +\Omega_0]$ corresponding to the spectral width of the laser pulse;
2) **this integral can be approximated with arbitrary accuracy using Gaussian quadrature formulae of sufficiently high order;**
3) the number of the cells in SLM can be as large as needed.

It is not clear now how the scheme described above can be extended to the case of neurons with multiple inputs $(d > 2)$. Nevertheless, it can be supposed, that as NMR can be considered as the first ready quantum computing technique, the already developed experimental methods of femtochemistry can be suitable for implementing quantum neural processing. Moreover, it has been already argued [19] that complex laser pulses can be used to produce single integrated instructions which can replace many consequent few-bit operations of quantum computers (the Complex Instruction Set quantum Computing (CISqC) technology [19]). So, quantum neural systems can be considered as a *flexible* tools able to optimize quantum computations by forming integrated instructions. It is reasonable to stress the importance of the flexibility of quantum neural processing. Ong, Huang, Tarn and Clark have already proved non-constructive existence theorem for the complete controllability of a class of quantum mechanical systems [20]. Their theorem states that it is possible to control systems with a discrete spectrum so that a desired final state will be fully occupied in a finite number of steps. As a rule, quantum control problems have many (even infinite) solutions and some additional restrictions are used to choose the best one [21]. This non-uniqueness is a clear indirect indication of the possibility of the development of a flexible neuron-like scheme able to perform arbitrary control (if it is universal) depending on the form of input signals. If these signals can be received from the controlled quantum system (closed-loop control), then it will be possible to perform arbitrary complex dynamical control of the process using a once-trained quantum neural system. Analogous feedback loop schemes are also under consideration in other fields of quantum computing. They can be used for continuous qubit purification [22]. This procedure can be quite important because quantum algorithms require some source of fresh qubits with defined initial states. From this point of view, quantum neural schemes should also be tested for possible application as flexible and trainable tools for temporal control of these desired pure quantum states.

**Conclusion**

We investigate both theoretically and by simulations some variants of a quantum neuron. It has been shown, in contrast to classical analog, some variants of single quantum neuron are able to perform approximation of any continuous function of many variables. It is also argued, that experimental methods of nonlinear optics and femtochemistry applied to coherent quantum control of multiphoton transitions can be suitable for implementation of quantum neuron processing system. Further comprehensive investigations of this point are needed in order to confirm or reject this suggestion.

**Acknowledgements**

We would like to thank G.D. Doolen, V.A. Mishchenko, A.L. Lunin, and A.V. Nifanova for useful discussions. The work was supported by the Department of Energy (DOE) under the contract W-7405-ENG-36. The work of GPB was partly supported by the National Security Agency (NSA), and by the Advanced Research and Development Activity (ARDA). AAE and AGK acknowledge the support of Russian Ministry of Industry and Science. AAE gratefully acknowledges the hospitality of the Theoretical Division and the Center for Nonlinear Studies of the Los Alamos National Laboratory.

Figures captions

Fig.1 *Left*: An experimental setup consisting of multi-barrier, multiple-slit system with a source of quanta (S) and a detector (D). One particle's trajectory corresponds to one universe in Everett's interpretation of quantum mechanics. The vector of refractivity indexes, **n,** is the input to the system. The trajectory's arms define weights and amplitudes at the detector with a complex-valued output of the single neuron. *Right*: Quantum neuron realized by the experimental setup.

Fig 2. Experimental realization of XOR function using collaborating quantum neurons in two different universes.

Fig 3. Two variants of the continuous mapping performed by a single barrier with 4 slits, trained to realize the $P = AND(n_1, n_2)$ function $(n_{1,2} \in [1, 1.67])$. A good generalization (smooth mapping) characterizes the mapping shown at the top (in this case for trained system $\Delta n \Delta l \leq \lambda/2$), while a poor generalization (bottom) is expressed as a wavy surface ($\Delta n \Delta l \cong 4\lambda$).

Fig 4. The waveguide model having two barriers with two slits each represents quantum neuron with two inputs and one bias value. The trajectories in the first universe is repeated 4 times $(k_1 = 4)$, while those in the second universe 2 times $(k_2 = 2)$. Lines of the same thickness correspond to the same value of refractory index.

Fig 5. The multiphoton absorption neural model. A programmable one-dimensional spatial light modulator (SLM) tunes both the spectral amplitude and the phase of the Ti: sapphire laser, spatially separated with diffraction grating. Phase tuning is performed in a factorized manner. One of the photons traversing the upper path with frequency lower than $\omega_0/2$ is tuned by the factor $x_1$, and other photon traverses the lower path with complementary frequency $\omega_2 = \omega_0 - \omega_1 > \omega_0/2$ and is tuned by the factor $x_2$. This pair of photons forms one of many possible interfering alternatives for 2-photon absorption by Cs atoms. The level of the occupancy of excited state of Cs monitored through fluorescence gives the desired output. Trained neuron parameters are stored in the memory and are used to control the SLM.

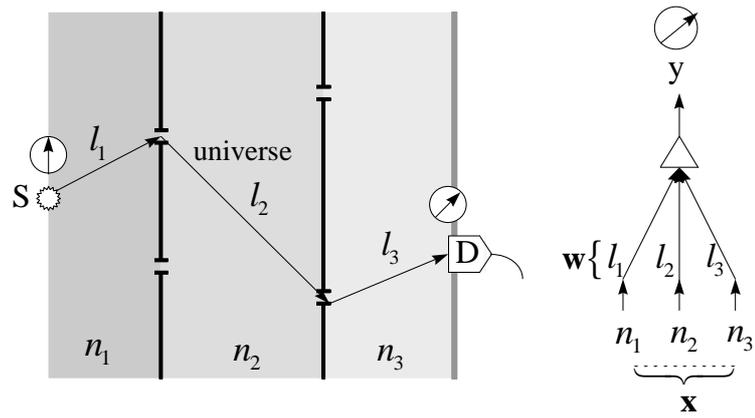

**Fig.1**

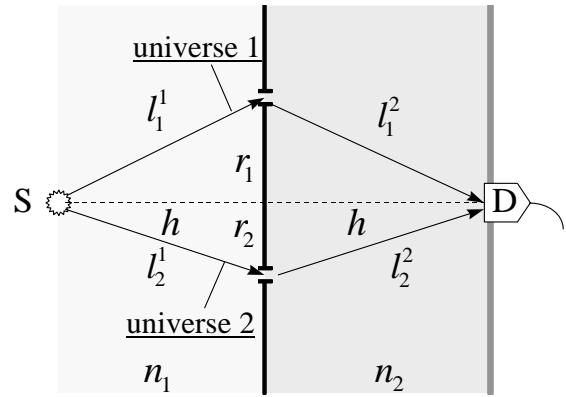

**Fig 2**.

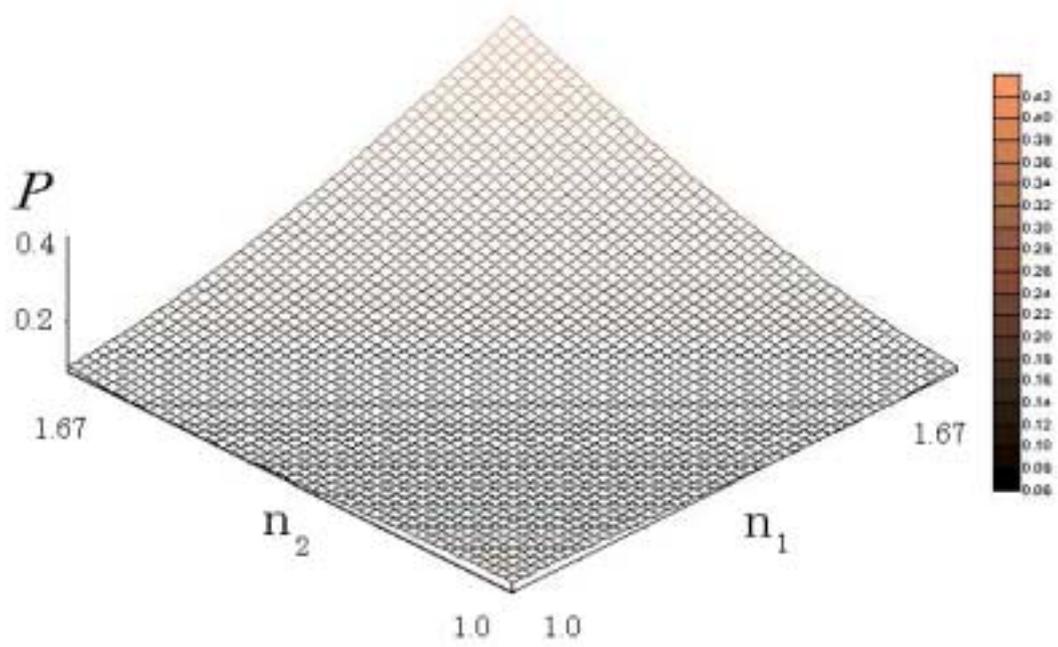
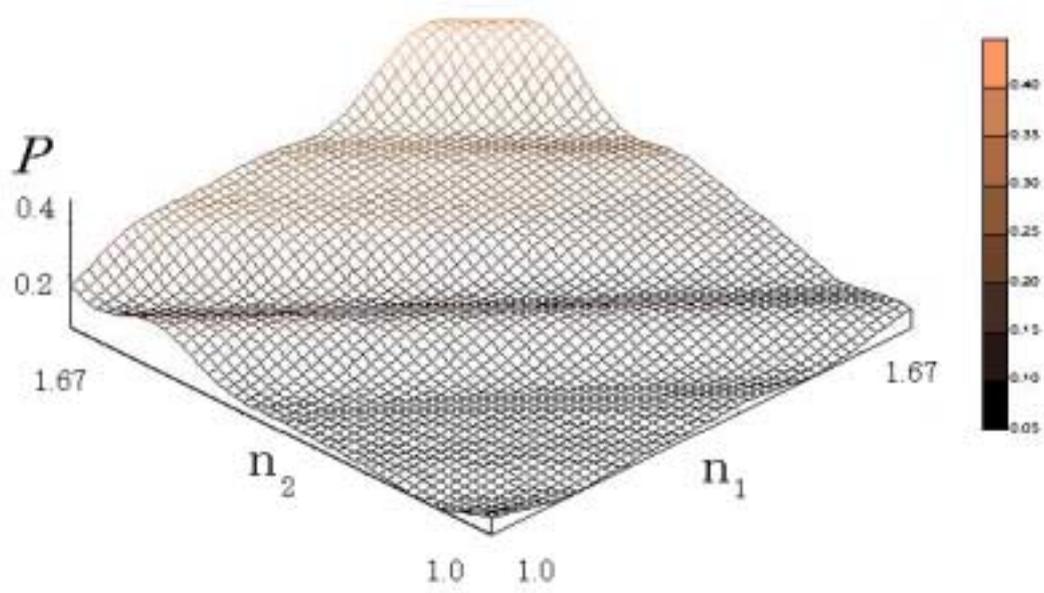

**Fig. 3**

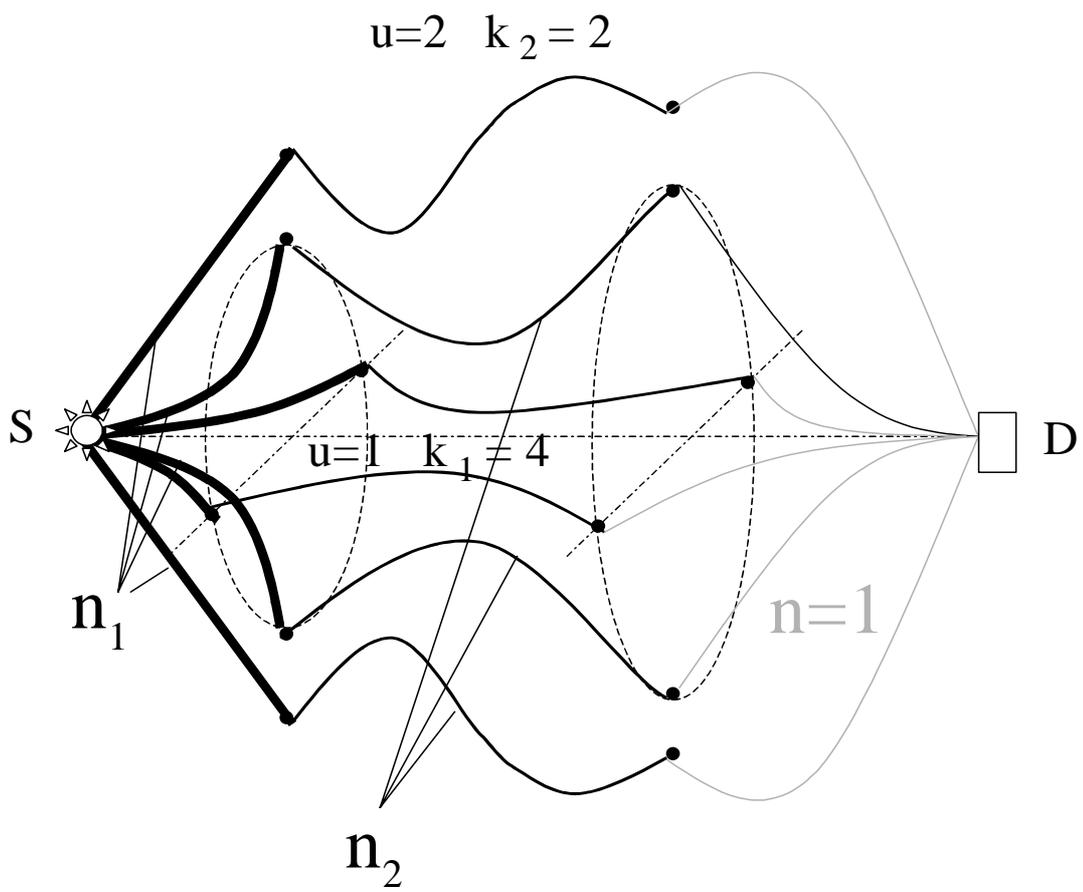

Fig 4.

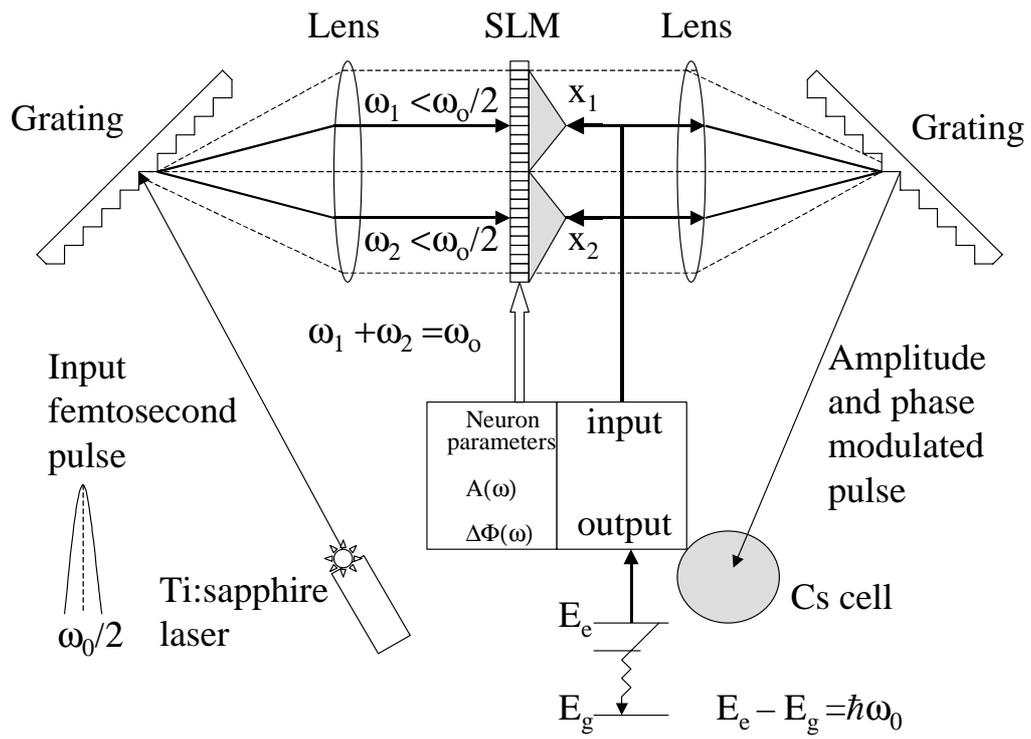

Fig.5